\documentclass[preprintnumbers,article,amsmath,amssymb,floatfix,10pt,prd,twocolumn,superscriptaddress,nofootinbib]{revtex4-2}
\usepackage{bm}
\usepackage{amsfonts}
\usepackage{latexsym}
\usepackage[utf8]{inputenc}
\usepackage{graphicx}
\usepackage{mathpazo}
\usepackage[]{mdframed}
\usepackage{textcomp}
\usepackage{float}
\usepackage{booktabs}
\usepackage{dcolumn}
\usepackage{ragged2e}
\usepackage{hyperref}
\usepackage{tensor}
\usepackage[version=3]{mhchem}
\hypersetup{colorlinks,citecolor=blue}
\hypersetup{colorlinks=true,linkcolor=red,filecolor=magenta,    urlcolor=blue}
\usepackage{xcolor}
\usepackage{enumitem}
\usepackage{orcidlink}
\usepackage{epsfig}
\usepackage[caption=false]{subfig}
\usepackage{commath}
\usepackage{array}
\usepackage{multirow}
\usepackage{tabularx}

\renewcommand{\arraystretch}{1.5}

\def\jnl@style{\it}
\def\aaref@jnl#1{{\jnl@style#1}}

\def\aaref@jnl#1{{\jnl@style#1}}

\def\aj{\aaref@jnl{AJ}}                   
\def\apj{\aaref@jnl{ApJ}}                 
\def\apjl{\aaref@jnl{ApJ}}                
\def\apjs{\aaref@jnl{ApJS}}               
\def\apss{\aaref@jnl{Ap\&SS}}             
\def\aap{\aaref@jnl{A\&A}}                
\def\aapr{\aaref@jnl{A\&A~Rev.}}          
\def\aaps{\aaref@jnl{A\&AS}}              
\def\mnras{\aaref@jnl{Mon.~Not.~Roy.~Astron.~Soc.}}             
\def\prd{\aaref@jnl{Phys.~Rev.~D}}        
\def\prc{\aaref@jnl{Phys.~Rev.~C}}  
\def\prl{\aaref@jnl{Phys.~Rev.~Lett.}}    
\def\qjras{\aaref@jnl{QJRAS}}             
\def\skytel{\aaref@jnl{S\&T}}             
\def\ssr{\aaref@jnl{Space~Sci.~Rev.}}     
\def\zap{\aaref@jnl{ZAp}}                 
\def\nat{\aaref@jnl{Nature}}              
\def\aplett{\aaref@jnl{Astrophys.~Lett.}} 
\def\apspr{\aaref@jnl{Astrophys.~Space~Phys.~Res.}} 
\def\physrep{\aaref@jnl{Phys.~Rep.}}      
\def\physscr{\aaref@jnl{Phys.~Scr}}       
\def\commat{\aaref@jnl{Comm.~Math.~Phys.}}              
\def\science{\aaref@jnl{Science}}               
\def\cqg{\aaref@jnl{Classical Quant.~Grav.}}            
\def\jpcs{\aaref@jnl{JPCS}}                                     
\def\ijmpd{\aaref@jnl{Int.~J.~Mod.~Phys.~D}}                    
\def\grg{\aaref@jnl{Gen.~Relat.~Gravit.}}               
\def\rpp{\aaref@jnl{Rep.~Prog.~Phys.}}          
\def\npa{\aaref@jnl{Nucl.~Phys.~A}}        
\def\lrr{\aaref@jnl{Living Rev.~Rel.}}                   
\def\jcap{\aaref@jnl{J.~Cosmology Astropart.~Phys.}}    
\def\rmp{\aaref@jnl{Rev.~Mod.~Phys.}}   
\def\epjc{\aaref@jnl{Eur.~Phys.~J.~C}} 
\def\plb{\aaref@jnl{~Phy.~Lett.~B}} 
\def\mpla{\aaref@jnl{Mod.~Phy.~Lett.~A}} 
\def\arxiv{\aaref@jnl{arxiv.org}}

\date{\today}

\begin{document}

\title{Beyond the Cosmological Constant: Breaking the Geometric Degeneracy of \texorpdfstring{$ f(Q) $}{Lg} cosmology via Redshift-Space Distortions}

\author{Ameya Kolhatkar\orcidlink{0000-0002-9553-1220}}
\email{kolhatkarameya1996@gmail.com}
\affiliation{Department of Mathematics, Birla Institute of Technology and Science, Pilani, Hyderabad Campus, Jawahar Nagar, Kapra Mandal, Medchal District, Telangana 500078, India}

\author{P.K. Sahoo\orcidlink{0000-0003-2130-8832}}
\email{pksahoo@hyderabad.bits-pilani.ac.in}
\affiliation{Department of Mathematics, Birla Institute of Technology and Science, Pilani, Hyderabad Campus, Jawahar Nagar, Kapra Mandal, Medchal District, Telangana 500078, India}

\begin{abstract}

We present a rigorous theoretical and observational analysis of the Hybrid $ f(Q) $ class of models by including the late-time modifying $ 1/Q $ term. After deriving strict viability conditions from the analytical expansion history, we show that preserving early-universe structure formation dictates that the linear coupling be exactly unity. This fixes the background of the Hybrid model into a geometric degeneracy with $ \Lambda $CDM which is confirmed explicitly through MCMC analysis with the latest background-only probes. The physical novelty of this model is manifest in the perturbation sector, where the geometric coupling breaks the background degeneracy and induces a late-time suppression of the effective gravitational constant $ G_{eff} < G_N $. Consequently, the inclusion of RSD data reveals an amplitude compensation mechanism, by which the matching of the signature $ f\sigma_8 $ of the data causes the clustering amplitude $ \sigma_8 $ to inflate under weaker gravity. Statistical model comparison through AIC/DIC demonstrates that incorporating growth data yields a moderate to weak preference for the Hybrid model keeping the background cosmology intact. This provides a physically bounded alternative to $ \Lambda $CDM with a falsifiable signature in the large scale structure, directly testable by the next generation of galaxy surveys.
    
\end{abstract}

\maketitle
\section{Introduction}
The $ \Lambda $CDM model provides a minimal, statistically robust description of our Universe consistent with modern observations ranging from Big Bang Nucleosynthesis (BBN) \cite{Cyburt:2015mya} to the Cosmic Microwave Background (CMB) \cite{Planck:2018vyg} to the Large-scale Structure (LSS) \cite{eBOSS:2020yzd} of late-time. The cosmological constant $ \Lambda $, which is the models answer to the observed cosmic acceleration \cite{SupernovaSearchTeam:1998fmf, SupernovaCosmologyProject:1998vns}, suffers from the catastrophic vacuum energy fine-tuning problem \cite{Weinberg:1988cp}. This is one of the many gaps in the conventionally formulated concordance model, which together motivate the search for a more consistent picture of the Universe. One such picture arises from the geometric modifications in the form of Symmetric Teleparallel Equivalent of General Relativity (STEGR) \cite{Nester:1998mp, BeltranJimenez:2017tkd, BeltranJimenez:2019esp, BeltranJimenez:2019tme, Heisenberg:2023lru}, whose models intend to explain the late-time acceleration through pure geometry without a need for a tunable vacuum field. In this work, we focus on the extended STEGR or the $ f(Q) $ theory which explains gravitational effects entirely through the non-metricity tensor $ Q_{\alpha\beta\gamma} $ and scalar $ Q $ of a flat, torsion-free connection. The $ f(Q) $ theory has generated a significant amount of attention \cite{BeltranJimenez:2018vdo, Lu:2019hra, BeltranJimenez:2019tme, Mandal:2020buf, Mandal:2020lyq, Barros:2020bgg, Lin:2021uqa, Dimakis:2021gby, Hassan:2021egb, Atayde:2021pgb, Frusciante:2021sio, Hassan:2022hcb, Sokoliuk:2023ccw, Gadbail:2023loj, Atayde:2023aoj, Mishra:2024rut, Capozziello:2024vix, Capozziello:2024jir, Mishra:2024shg, Boiza:2025xpn, Kolhatkar:2025ubm, Capozziello:2025hyw, Kavya:2025vsj, Li:2025msm, Kolhatkar:2026ixl} due to the strictly second-order field equations, avoiding the Ostrogradsky instabilities that plague higher-order metric theories.\\

In the coincident gauge the non-metricity scalar for the spatially flat FLRW metric evaluates to $ Q=6H^2 $ where $ H\equiv H(z) $ is the Hubble parameter. Realizing that $ H $ increases monotonically with increasing redshift $ z $, one can patch together a simple non-trivial model of the form $ f(Q)\sim Q + 1/Q $ which only modifies the late-time dynamics. It is clear that this gives back GR like behavior at early times. We call this the Hybrid model which has been previously explored in the literature \cite{BeltranJimenez:2019tme, Rana:2024mst, Kolhatkar:2024oyy, Atayde:2026upv}.\\

In this work, we wish to express the importance of deriving analytical viability conditions before injecting the model into a MCMC sampler, especially for non-trivial cosmologies. Without prior knowledge of the parameter boundaries and the forbidden regions, a sampler can enter unphysical regions and give rise to instabilities. Hence, we start by deriving the mathematically viable constraints for the Hybrid model and show that in order to preserve high redshift behavior, the linear coupling has to be exactly unity. This causes the early time behavior of the model to become degenerate with $ \Lambda $CDM. The late-time probes including Cosmic Chronometers, Type Ia Supernovae and Baryon Acoustic Oscillations do not break this background degeneracy. We thus include growth data invoking the linear perturbation sector, and find that the Hybrid coupling suppresses the effective Gravitational constant $ G_{eff} $ at late-times. However, the clustering amplitude $ \sigma_8 $ is increased in order to accommodate the observed values $ f\sigma_8 $.\\

This manuscript is organized as follows: \autoref{section_fQIntro} gives the convention and equations for $ f(Q) $ gravity followed by \autoref{section_analyticalconstraints} that outline analytical constraints as well as model branch selection. We also show the linear coupling, $ \alpha_1=1 $ here. \autoref{section_obsandmeth} then mentions the observational datasets and the statistical methodology used. \autoref{section_results} demonstrates the results of background and perturbation analyses. Finally, we conclude our manuscript in \autoref{section_conclusion} and discuss the feasibility of the model in light of the upcoming next-generation surveys.
\section{\texorpdfstring{$ f(Q) $}{Lg} Cosmology and the Hybrid Model}\label{section_fQIntro}
\subsection{\texorpdfstring{$ f(Q) $}{Lg} Cosmology}
The $ f(Q) $ gravitational framework is an extension of the STEGR formalism by generalizing the geometric action. We begin with a general affine connection $ \tensor{A}{^\alpha_{\mu\nu}} $ that admits splitting into three components\cite{Ortin:2015hya}
\begin{equation}\label{eq_connectionsplit}
    \tensor{A}{^\alpha_{\mu\nu}} = \tensor{\hat{A}}{^\alpha_{\mu\nu}} + \tensor{K}{^\alpha_{\mu\nu}} + \tensor{L}{^\alpha_{\mu\nu}}\;.
\end{equation}
The other two components along with the Levi-Civita connection are the contortion tensor $ \tensor{K}{^\alpha_{\mu\nu}} $ and the disformation tensor $ \tensor{L}{^\alpha_{\mu\nu}} $ which are in turn built from the torsion tensor $ \tensor{T}{^\alpha_{\mu\nu}} = 2\tensor{A}{^\alpha_{[\mu\nu]}} $ and the non-metricity tensor $ Q_{\alpha\mu\nu}=\nabla_\alpha g_{\mu\nu} $ as follows
\begin{gather}
    \tensor{\hat{A}}{^\alpha_{\mu\nu}}=\frac{1}{2}g^{\alpha\lambda}\left( 2\partial_{(\mu}g_{\nu)\lambda} - \partial_\lambda g_{\mu\nu} \right) \;,\\
    \tensor{K}{^\alpha_{\mu\nu}}=\frac{1}{2}\tensor{T}{^\alpha_{\mu\nu}}+\tensor{T}{_{(\mu}^\alpha_{\nu)}}\;,\\
    \tensor{L}{^\alpha_{\mu\nu}}=\frac{1}{2}\tensor{Q}{^\alpha_{\mu\nu}}-\tensor{Q}{_{(\mu}^\alpha_{\nu)}}\;.
\end{gather}
By imposing a torsion-free, flat spacetime through $ \tensor{T}{^\alpha_{\mu\nu}} = 0 $ and $ \tensor{R}{^\alpha_{\mu\beta\nu}} = 0 $ we obtain the relation $ \hat{R} = Q + \nabla_\alpha(Q^\alpha - \Tilde{Q}^\alpha) $. The non-metricity scalar is computed using $ Q = Q_{\alpha\mu\nu}P^{\alpha\mu\nu} $ with the superpotential defined as
\begin{equation}
    4\tensor{P}{^\alpha_{\mu\nu}} = -2\tensor{L}{^\alpha_{\mu\nu}} + (Q^\alpha - \Tilde{Q}^\alpha)g_{\mu\nu} - \delta^\alpha_{(\mu}Q_{\nu)}
\end{equation}
and the independent traces of the non-metricity tensor given by $ Q_\alpha=g^{\mu\nu}Q_{\alpha\mu\nu} $ and $ \Tilde{Q}^\alpha=g^{\mu\nu}Q_{\mu\nu\alpha} $. The action of $ f(Q) $ theory reads
\begin{equation}\label{eq_action}
    S = \int d^4x\sqrt{-g}\left( \frac{1}{2\kappa}f(Q) + \mathcal{L_M} \right)\;.
\end{equation}
Here, $ \kappa=8\pi G $ and $ \mathcal{L_M} $ is the Lagrangian corresponding to the matter fields. The field equations corresponding to \eqref{eq_action} are given by
\begin{multline}\label{eq_fQfieldequations}
    \frac{2}{\sqrt{-g}}\nabla_\alpha(\sqrt{-g}f_Q\tensor{P}{^\alpha_{\mu\nu}})-\frac{1}{2}g_{\mu\nu}f+\\
f_Q(P_{\mu\alpha\beta}\tensor{Q}{_\nu^{\alpha\beta}}-2Q_{\alpha\beta\mu}\tensor{P}{^{\alpha\beta}_\nu})=\kappa T_{\mu\nu}\;.
\end{multline}
We now adopt the coincident gauge, in which the flat spacetime without torsion allows a parameterization of the connection as $ \tensor{A}{^\alpha_{\mu\nu}} = \frac{\partial x^\alpha}{\partial\xi^\beta}\partial_\mu\partial_\nu\xi^\beta $. Thus, for the ``gauge" choice $ \xi^\alpha=x^\alpha $, the connection can be set to zero globally, causing the tangent space of $ \xi^\alpha $ to coincide with the spacetime. Hence, the name coincident gauge. Furthermore, we choose the spatially flat FLRW metric $ g_{\mu\nu} = \text{diag}(-1, a^2(t), a^2(t), a^2(t)) $ which results in a simple form for the non-metricity scalar $ Q=6H^2 $. The field equations \eqref{eq_fQfieldequations} become the modified Friedman equations
\begin{gather}
    3H^2 = \kappa (\rho + \rho_D)\label{eq_modified_friedmann1}\\
    3H^2 + 2\dot{H} = -\kappa(p+p_D)\label{eq_modified_friedmann2}
\end{gather}
where the density and pressure corresponding to the effective dark energy such as fluid is expressed in terms of the geometry set by the choice of $ f(Q) $ 
\begin{gather}
    2\kappa\rho_D = Q(1-2f_Q) + f\\
    2\kappa p_D = 4\dot{H}(f_Q - 1) - f + Q(8f_{QQ}\dot{H} + 2f_Q - 1)
\end{gather}
While the above equations govern the background expansion, the geometric modification alters the perturbation sector. As explained in \cite{BeltranJimenez:2019tme}, under the quasi-static approximation, the effective gravitational constant scales by $ \mu(z)=1/f_Q $. This allows for a quantitative study of the growth of structure in the linear perturbation sector. 
\subsection{The Hybrid \texorpdfstring{$ f(Q) $}{Lg} class}
The Hybrid model below is written down as a seemingly 3 parameter model with linear and inverse terms of $ Q $, along with the provision of the cosmological constant like term\footnote{$ Q_0 $ is the present day value of $ Q $}. 
\begin{equation}\label{eq_model_form}
    f(Q) = \alpha_1 Q + \alpha_2 Q_0 + \alpha_3\frac{Q_0^2}{Q}\;.
\end{equation}
The inverse term dominates only at late-times, whereas at the early epochs, the Lagrangian essentially reduces to $ \Lambda $CDM. Although this model is phenomenologically attractive, the coupling coefficients cannot be arbitrarily chosen. The next section is devoted to deriving analytical bounds on the apparent free parameters. 
\section{Analytical validity constraints}\label{section_analyticalconstraints}
Inserting \eqref{eq_model_form} in \eqref{eq_modified_friedmann1} and \eqref{eq_modified_friedmann2} yields two analytical solutions for $ H(z) $
\begin{equation}
    H_{\pm}(z) = H_0\sqrt{\frac{\Tilde{\Omega}(z) \pm \sqrt{\Tilde{\Omega}^2(z) + 12\alpha_1\alpha_3}}{2\alpha_1}}
\end{equation}
with $ \Tilde{\Omega}(z) = \Omega(z) + \alpha_2 = \Omega_{m0}(1+z)^3 + \Omega_{r0}(1+z)^4 + \alpha_2 $. The present condition along with $ \alpha_1\neq0 $ imposes the constraint $ 3\alpha_3 = \alpha_1 - \Tilde{\Omega}_0 $ that eliminates one of the three couplings:
\begin{equation}
    H_{\pm}(z) = H_0\sqrt{\frac{\Tilde{\Omega}(z) \pm \sqrt{\Tilde{\Omega}^2(z) + 4\alpha_1(\alpha_1-\Tilde{\Omega}_0)}}{2\alpha_1}}\;.
\end{equation}
It should be noted that $ \alpha_1>0 $ to avoid kinetic instabilities since the linear term dominates at early times. Furthermore, since $ \alpha_2 $ acts as the cosmological constant at high redshifts, we restrict it to non-negative values -- $ \alpha_2\geq0 $. With these rudimentary bounds established, the negative branch is immediately discarded. At high redshifts, the nested square root reduces to $ \sqrt{\Tilde{\Omega}^2(z) + 4\alpha_1(\alpha_1-\Tilde{\Omega}_0)} \sim \Omega(z) $. This causes $ H_-(z)\sim0 $ for the far past suppressing the early expansion rate. The physically viable expansion history is thus governed by the positive branch
\begin{equation}\label{eq_hubble_final}
    H_+(z)\equiv H(z) = H_0\sqrt{\frac{\Tilde{\Omega}(z) + \sqrt{\Tilde{\Omega}^2(z) + 4\alpha_1(\alpha_1-\Tilde{\Omega}_0)}}{2\alpha_1}}
\end{equation}
The first constraint of validity on the physical branch is obtained from the positivity of the argument under the nested square root; $ \Tilde{\Omega}^2(z) + 4\alpha_1(\alpha_1-\Tilde{\Omega}_0) \geq 0 $. We observe that since $ \Omega(z) $ and therefore $ \Tilde{\Omega}(z) $ increases monotonically with increasing $ z $, the most stringent condition will be obtained at $ z=z_{min}=-1 $ regardless of the sign and magnitude of $ \alpha_2 $. This ensures inclusion of the asymptotic future. Since $ \Omega(z=-1)=0 $, we have 
\begin{gather}
    \alpha_2^2 + 4\alpha_1(\alpha_1-\alpha_2-\Omega_0)\geq 0\\
    \therefore\;(\alpha_2-2\alpha_1+2\sqrt{\alpha_1\Omega_0})(\alpha_2-2\alpha_1-2\sqrt{\alpha_1\Omega_0})\geq0
\end{gather}
Thus, our first constraint of viability for the positive variant is given by
\begin{equation}
    v_{+,1}\;:\;\alpha_2\in\left[0, \alpha_{2,-}\right] \cup \left[\alpha_{2,+},+\infty\right)
\end{equation}
where $ \alpha_{2,\pm}=2(\alpha_1 \pm \sqrt{\alpha_1\Omega_0}) $. The second constraint equation is obtained by demanding the positivity of the argument under the primary square root retaining for all $ z\geq-1 $. Given that $ \alpha_2\geq0 $, this quantity is always positive.\\

\begin{figure}[ht]
    \centering
    \includegraphics[width=0.5\textwidth]{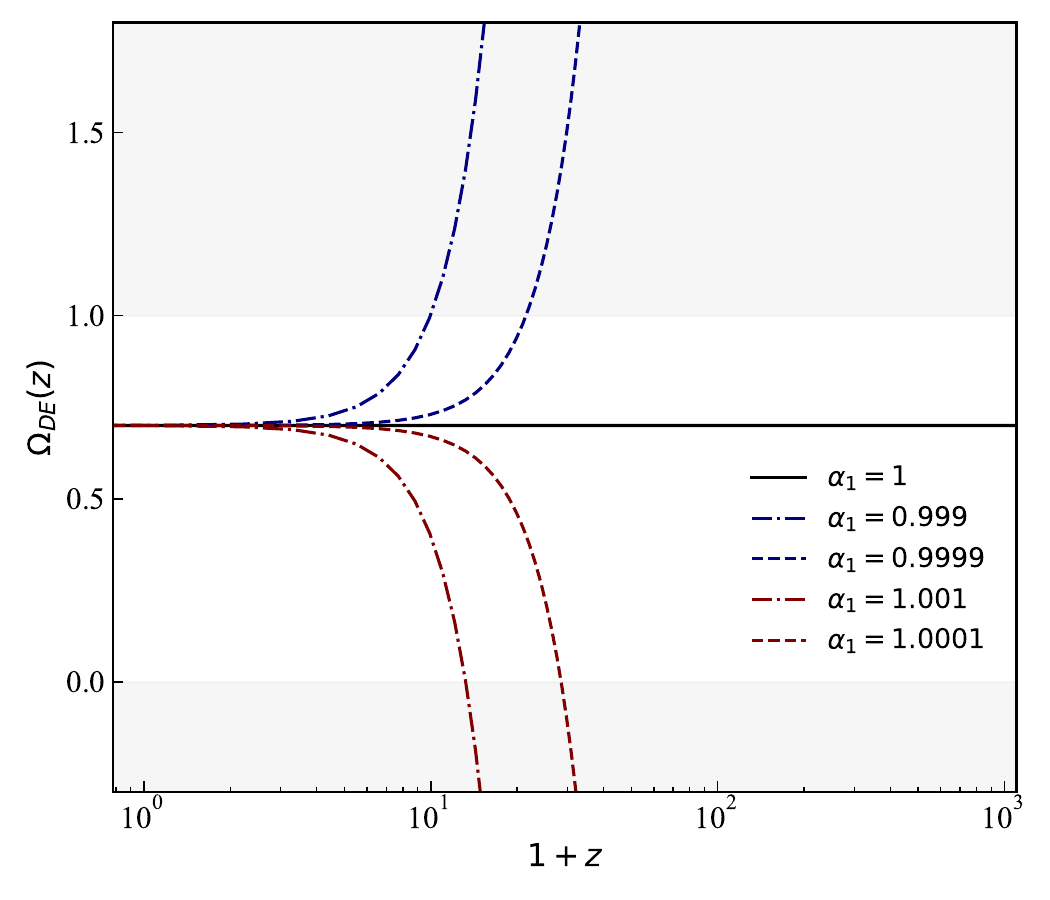}
    \caption{Evolution of the effective dark energy density. Small deviations from $\alpha_1 = 1$ trigger a catastrophic Early Dark Energy domination at high redshifts, mathematically necessitating $\alpha_1 = 1$ to preserve physical early-universe cosmology.}
    \label{figure_omegadeplot}
\end{figure}

Equation \eqref{eq_hubble_final} can be cast into a form resembling a spatially flat Universe with an evolving Dark Energy given 
\begin{multline}\label{eq_effective_DE}
    2\alpha_1\Omega_{DE}(z) = 2\alpha_1\alpha_2 + (1-2\alpha_1)\Tilde{\Omega}(z)\\ + \sqrt{\Tilde{\Omega}^2(z) + 4\alpha_1(\alpha_1-\Tilde{\Omega}_0)}\;. 
\end{multline}
In deep matter and radiation eras, this effective dark energy density reduces to  
\begin{equation}
    \Omega_{DE}(z\gg1) = \frac{1}{2\alpha_1}\left( 2\alpha_1\alpha_2 + 2(1-\alpha_1)\Tilde{\Omega}(z) \right)\;.
\end{equation}
This limit reveals a catastrophic physical consequence. Any deviation from $ \alpha_1 = 1 $ will be enormously amplified by the $ (1+z)^3 $ or $ (1+z)^4 $ scaling. As shown in \autoref{figure_omegadeplot}, even a minor change in $ 0.1\% $ induces Early Dark Energy that destroys the precise conditions required for the formation of acoustic peaks of the BBN and CMB. Thus, in order to preserve the formation of the structure tested in the observation, we are mathematically forced to set $ \alpha_1=1 $ throughout this analysis. This bound locks the background expansion history of the Hybrid model in a degeneracy with $ \Lambda $CDM, where $ \Omega_{DE}(z\gg1) = \alpha_2 $. This geometric degeneracy substantiates the fact that the late-time physical novelty of this model can only be probed through the linear perturbation sector.
\section{Observational Analysis and Statistical methodology}\label{section_obsandmeth}
\subsection{Observational Data}
Having established that the Hybrid model exhibits a background degeneracy with $ \Lambda $CDM, we divide our dataset combinations into two categories -- background only probes and background + growth. Refer to \autoref{table_priors} for a list of priors.
\begin{table}[]
    \centering
    \begin{tabular}{>{\centering\arraybackslash}m{8em}>{\centering\arraybackslash}m{8em}}
    \hline
        \textbf{Parameter} & \textbf{Priors} \\
        \hline
        $ H_0 $ & $ \mathcal{U}[\;50, 90\;] $\\
        $ \Omega_{m0} $ & $ \mathcal{U}[\;0.1, 0.9\;] $\\
        $ \mathcal{M} $ & $ \mathcal{U}[\;-19.6,-18.8\;] $\\
        $ \sigma_{80} $ & $ \mathcal{U}[\;0.6,1.2\;] $\\
        $ \alpha_2 $ & $ \mathcal{U}[\;0,5\;] $\\
        \hline
    \end{tabular}
    \caption{A list of all the priors used. $ \mathcal{U}[a,b] $ denotes a flat prior with the upper bound $ b $ and lower bound $ a $.}
    \label{table_priors}
\end{table}
\subsubsection{Background only probes}
This category employs two combinations - Base ( CC + SN + BAO) and Base + SH0ES. 
\begin{itemize}
    \item \textbf{\textit{Cosmic Chronometers - CC}} : We use 32 points of the Cosmic Chronometer $ H(z) $ measurements with 15 correlated and 17 uncorrelated data points as given in Table II of \cite{deCruzPerez:2024shj} that span the redshift range $ 0.070 < z < 1.965 $. These data points are obtained by relying on the differential ages of passively evolving, ultra massive early galaxies without assuming a baseline cosmology. The procedure for correlated measurements is carried out as explained in \texttt{\url{https://gitlab.com/mmoresco/CCcovariance}}.
    \item \textbf{\textit{Type Ia Supernovae - SN}} : We use the \texttt{Pantheon+} compilation \cite{Brout:2022vxf, Scolnic:2021amr} after excluding all points with $ z<0.01 $ corresponding to model dependent data points. This leaves us with 1590 out of the total 1701 points spanning $ 0.01016<z<2.26137 $. We consider the absolute magnitude $ \mathcal{M} $ to be a free parameter.
    \item \textbf{\textit{Baryon Acoustic Oscillations - BAO}} : We utilize the most recent Baryon Acoustic Oscillation data as in the second release by the DESI collaboration \cite{DESI:2025zgx}. This data set measures three quantity ratios - $ D_M/r_d $, $ D_H/r_d $ and $ D_V/r_d $, where $ r_d $ is the radius of the sound horizon at the drag epoch and $ D_M, D_H $ and $ D_V $ correspond to the transverse co-moving distance, the Hubble distance and the volume-averaged distance respectively. Since the present model does not alter early Universe physics, we fix $ r_d = 147.05 $ Mpc.
    \item \textbf{\textit{SH0ES $ H_0 $ prior}} : In order to assess the response of the model to the Hubble tension, we supplement the Base dataset with local $ H_0 $ prior from \cite{Riess:2019cxk}.
\end{itemize}
\subsubsection{Growth data}
This category also employs two dataset combinations - RBase and RBase + SH0ES where RBase is Base + RSD. 
\begin{itemize}
    \item \textbf{\textit{Redshift-Space Distortion - RSD}} : In order to explore the linear perturbation sector, we employ a compilation of growth data constraining $ f\sigma_8 $ present in Table II of \cite{Alestas:2022gcg}. Both the growth rate $ f(a) = \frac{dln\delta_m}{dlna} $ and the clustering amplitude $ \sigma_8(a) = \sigma_{80}\frac{\delta_m(a)}{\delta_m(1)} $ are derived by solving the matter overdensity ($ \delta_m(a) $) evolution equation
    \begin{equation}\label{eq_growth}
        \delta_m'' + \delta_m'\left( 2 + \frac{H'}{H} \right) - \frac{3}{2}\delta_m\Omega_m\mu = 0\;,
    \end{equation}
    where primes denote derivatives with respect to the e-folding number $ N = ln(a) $. The modification of the effective gravitational constant is captured through $ \mu = G_{eff} / G_N = 1/f_Q $. For the Hybrid model, $ f_Q = 1 - \alpha_3(Q_0/Q)^2  $, and the sign of $ \alpha_3 $ determines the growth suppression / amplification in late times. 
\end{itemize}

\subsection{Bayesian Inference and diagnostics}
We deploy a custom, parallelized Bayesian inference pipeline using \texttt{emcee} sampler \cite{Foreman_Mackey_2013} to rigorously explore the parameter space and \texttt{GetDist} \cite{Lewis:2019xzd} to visualize the posterior distribution. The convergence of our multi-chain runs has been ensured by $ \hat{R} - 1 < 0.01 $. We compare our model with $ \Lambda $CDM through the Akaike Information Criterion (AIC) and the Deviance Information Criterion (DIC) computed using the minimum $ \chi^2_{min} $ value as below
\begin{gather}
    AIC = \chi^2_{min} + 2k\\
    DIC = \chi^2_{min} + 2p\;.
\end{gather}
Here, $ k $ is the number of independent model parameters and $ p = \Bar{\chi}^2 - \chi^2_{min} $ with the bar indicating the average of all $ \chi^2 $s from the MCMC chains.

\section{Results}\label{section_results}
\begin{figure}
    \centering
    \includegraphics[width=0.5\textwidth]{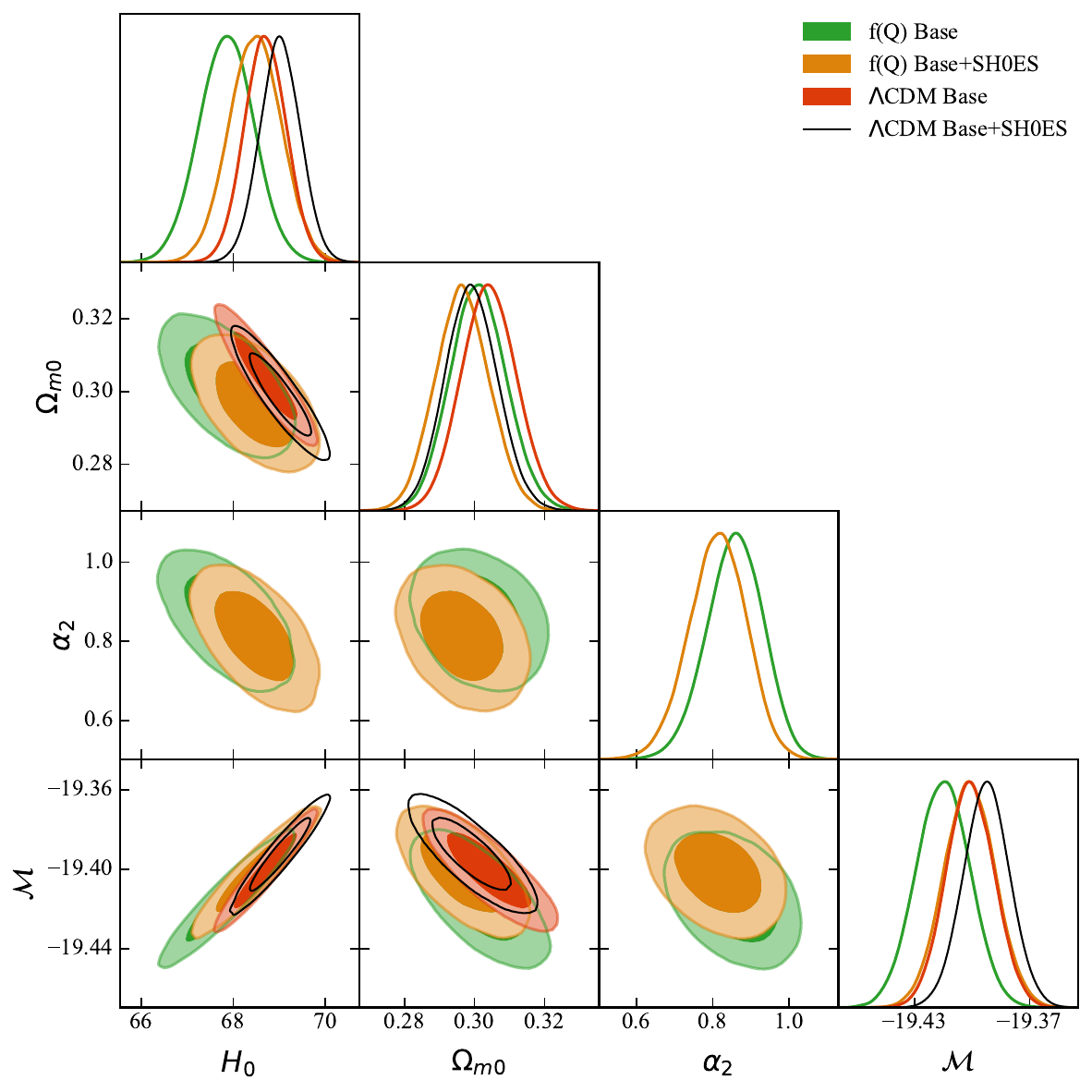}
    \caption{Posterior contours of both the models under Background probes}
    \label{figure_corner_BG}
\end{figure}

\begin{figure}
    \centering
    \includegraphics[width=0.5\textwidth]{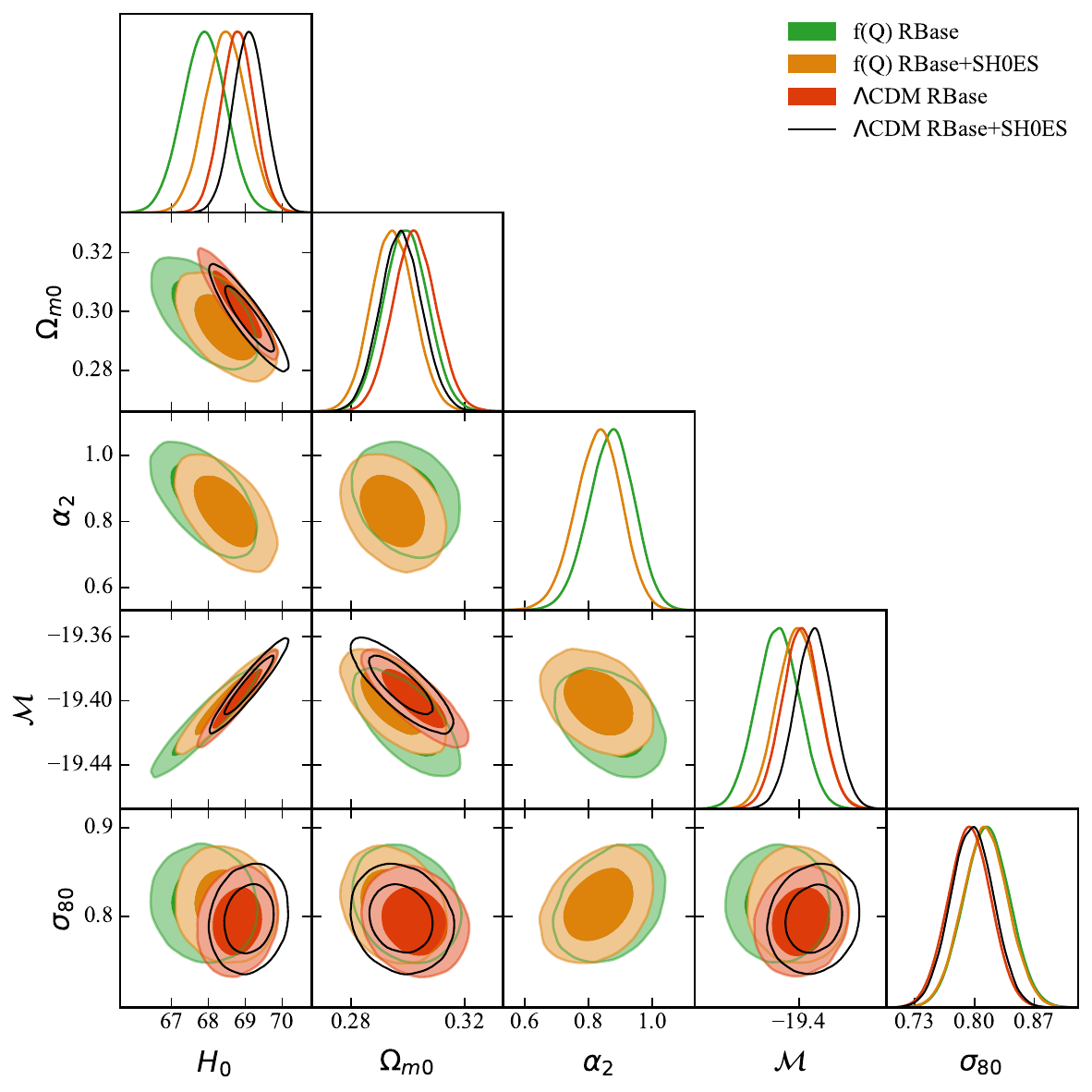}
    \caption{Posterior contours of both the models under Background+Growth}
    \label{figure_corner_BGG}
\end{figure}

In this section, we present the observational constraints derived from the MCMC analysis. The posteriors are visualized in \autoref{figure_corner_BG} and \autoref{figure_corner_BGG}, while the $ 1\sigma $ constraints on parameters with information criteria and present day values are summarized in \autoref{table_Result_Summary} and \autoref{table_presentdayvalues} respectively.
\subsection{Background Dynamics}
The primary objective of this section is verifying the ability of the analytically bounded Hybrid model in replicating the background expansion history. This is confirmed in \autoref{table_Result_Summary} where the cosmological parameters $ H_0 $ and $ \Omega_{m0} $ remain consistent with standard $ \Lambda $CDM predictions for all the dataset combinations. The information criteria however, uncover a statistical advantage of the Hybrid model. Despite the additional parameter $ \alpha_2 $, the base combination yields $ \Delta $AIC $ =-2.796 $ and $ \Delta $DIC $ =-2.785 $ demonstrating that the late-time $ 1/Q $ term provides a statistically superior fit to both SN and BAO as compared to a simple cosmological constant. Adding the SH0ES prior causes the model to accommodate a higher $ H_0 $ without breaking the background constraints.
\subsection{Perturbation Kinematics : Gravitational suppression and Amplitude Compensation}
This section establishes the novelty of the Hybrid $ f(Q) $ model through the linear perturbation sector. The parameter $ \alpha_2 $ is constrained at $ 0.871^{+0.075}_{-0.067} $ for RBase and $ 0.830\pm0.072 $ for RBase+SH0ES setting $ \alpha_3 < 0 $ and suppressing the effective gravitational constant $ \mu(z) < 1 $. The suppressed gravity gives way to an immediate consequence: the amplitude compensation mechanism. Since structure formation takes longer under weaker gravity, the Hybrid model now compensates for this by inflating the late-time clustering amplitude $ \sigma_{80} $ from $ 0.794\pm0.025 $ to $ 0.814\pm0.027 $\footnote{Similar results are found for the RBase+SH0ES combination}. This compensation mechanism provides an exceptionally tight fit to the new RBase combination, with $ \Delta $AIC $ =-3.721 $ and $ \Delta $DIC $ =-3.764 $. Moreover, despite being in the statistically compatible range, the RBase+SH0ES combination still pulls the information criteria metrics down from $ -0.39 $ to $ -1.05 $. These results establish the positive statistical preference of the Hybrid $ f(Q) $ model. 
 \begin{table*} 
        \centering
        \renewcommand{\arraystretch}{2}
        \begin{tabular}{ >{\centering\arraybackslash}m{6em} 
        >{\centering\arraybackslash}m{6.5em} 
        >{\centering\arraybackslash}m{6.5em}  
        >{\centering\arraybackslash}m{7.5em}
        >{\centering\arraybackslash}m{7.5em}
        >{\centering\arraybackslash}m{6.5em} 
        >{\centering\arraybackslash}m{4em}
        >{\centering\arraybackslash}m{4em} }
        \hline
        \textbf{MODEL} & $ \bm{H_0} $ & $ \bm{\Omega_{m0}} $ & $ \bm{\mathcal{M}} $ & $ \bm{\sigma_{80}} $ & $ \bm{\alpha_2} $ & $ \bm{\Delta} $\textbf{AIC} & $ \bm{\Delta} $\textbf{DIC} \\
        \hline
        \multicolumn{8}{c}{\textsc{\textbf{Base}}} \\
        \hline
        $ \bm{\Lambda} $\textbf{CDM} & $ 68.70\pm0.46 $ & $ 0.304\pm0.008 $ & $ -19.40\pm0.013 $ & --------- & --------- & $ 0 $  & $ 0 $ \\
        
        \textbf{Hybrid} & $ 67.86\pm0.60 $ & $ 0.301\pm0.008 $ & $ -19.42\pm0.014 $ & --------- & $ 0.858^{+0.077}_{-0.069} $ & $ -2.796 $ & $ -2.785 $ \\
        \hline
        \multicolumn{8}{c}{\textsc{\textbf{Base+SH0ES}}} \\
        \hline
        $ \bm{\Lambda} $\textbf{CDM} & $ 69.02\pm0.44 $ & $ 0.299\pm0.008 $ & $ -19.39\pm0.012 $ & --------- & --------- & $ 0 $ & $ 0 $ \\
        
        \textbf{Hybrid} & $ 69.48\pm0.57 $ & $ 0.297\pm0.008 $ & $ -19.40\pm0.013 $ & --------- & $ 0.813\pm0.075 $ & $ -0.380 $ & $ -0.406 $ \\
        \hline
        \multicolumn{8}{c}{\textsc{\textbf{RBase}}} \\
        \hline
        $ \bm{\Lambda} $\textbf{CDM} & $ 68.79\pm0.44 $ & $ 0.302\pm0.008 $ & $ -19.40\pm0.012 $ & $ 0.794\pm0.025 $ & --------- & $ 0 $ & $ 0 $ \\
        
        \textbf{Hybrid} & $ 67.89\pm0.59 $ & $ 0.299\pm0.008 $ & $ -19.41\pm0.014 $ & $ 0.814\pm0.027 $ & $ 0.871^{+0.075}_{-0.067} $ & $ -3.721 $ & $ -3.764 $ \\
        \hline
        \multicolumn{8}{c}{\textsc{\textbf{RBase+SH0ES}}}\\
        \hline
        $ \bm{\Lambda} $\textbf{CDM} & $ 69.10\pm0.44 $ & $ 0.298\pm0.007 $ & $ -19.39\pm0.012 $ & $ 0.797\pm0.025 $  & --------- & $ 0 $ & $ 0 $ \\
        
        \textbf{Hybrid} & $ 68.48\pm0.56 $ & $ 0.295\pm0.008 $ & $ -19.40\pm0.013 $ & $ 0.813\pm0.027 $ & $ 0.830\pm0.072 $ & $ -1.056 $ & $ -1.160 $ \\
        \hline
        \end{tabular}
        \caption{Summary of MCMC post-analysis}
        \label{table_Result_Summary}
    \end{table*}
\begin{table*}
    \centering
    \begin{tabular}{>{\centering\arraybackslash}m{8em}>{\centering\arraybackslash}m{10em}>{\centering\arraybackslash}m{8em}>{\centering\arraybackslash}m{8em}
    >{\centering\arraybackslash}m{8em}
    >{\centering\arraybackslash}m{8em}}
    \hline
        \textbf{Model} & \textbf{Combination} & $ \bm{q_0} $ & $ \bm{z_t} $ & $ \bm{\omega_{eff0}} $ & $ \bm{S_8} $ \\
        \hline
        \multirow{5}{4em}{$ \Lambda $CDM} & Base & $ -0.544 $ & $ 0.659 $ & $ -0.696 $ & ------\\
        & Base+SH0ES & $ -0.551 $ & $ 0.673 $ & $ -0.701 $ & ------\\
        & RBase & $ -0.547 $ & $ 0.665 $ & $ -0.698 $ & $ 0.796 $\\
        & RBase+SH0ES & $ -0.553 $ & $ 0.678 $ & $ -0.702 $ & $ 0.794 $\\
        \hline
        \multirow{5}{4em}{$ f(Q) $} & Base & $ -0.463 $ & $ 0.678 $ & $ -0.642 $ & ------\\
        & Base+SH0ES & $ -0.499 $ & $ 0.686 $ & $ -0.666 $ & ------\\
        & RBase & $ -0.459 $ & $ 0.687 $ & $ -0.640 $ & $ 0.813 $\\
        & RBase+SH0ES & $ -0.494 $ & $ 0.693 $ & $ -0.663 $ & $ 0.806 $\\
        \hline
    \end{tabular}
    \caption{Summary of present day cosmological parameters.}
    \label{table_presentdayvalues}
\end{table*}
\subsection{Reconstructed Cosmological Evolution}
\begin{figure}
    \centering
    \includegraphics[width=0.5\textwidth]{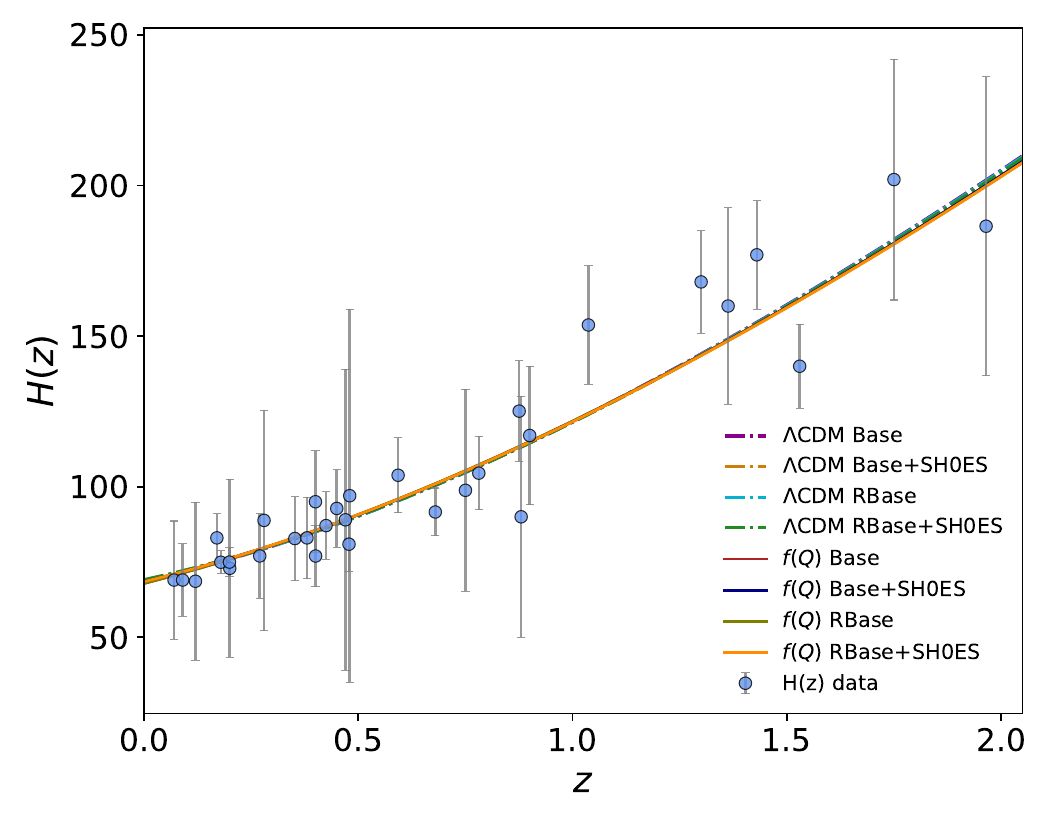}
    \caption{The reconstructed Hubble parameter.}
    \label{figure_Hz}
\end{figure}

\begin{figure}
    \centering
    \includegraphics[width=0.5\textwidth]{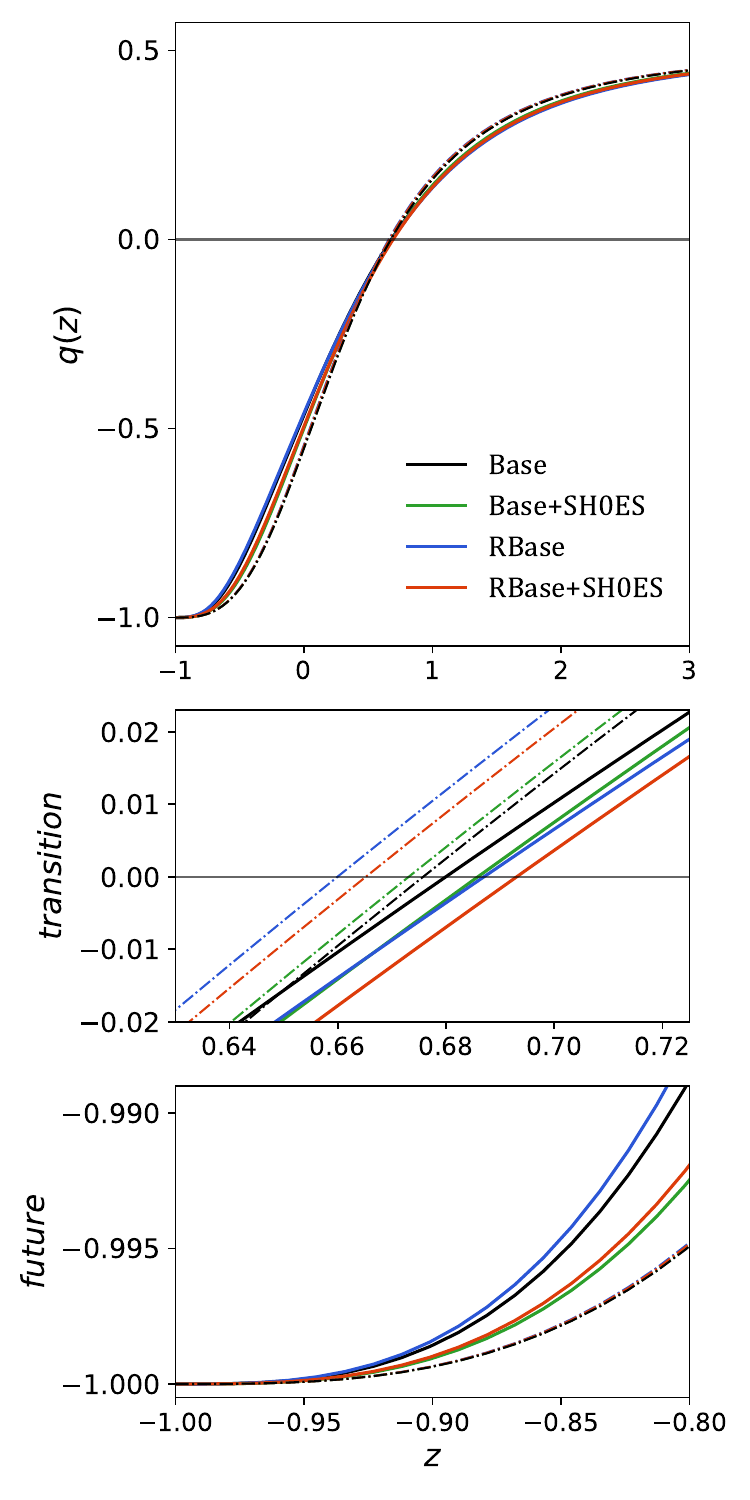}
    \caption{Deceleration parameter $ q(z) $. The solid lines represent $ f(Q) $ model while the dot-dashed lines represent $ \Lambda $CDM. Each color corresponds to the same combination of datasets.}
    \label{figure_deceleration}
\end{figure}

\begin{figure}
    \centering
    \includegraphics[width=0.5\textwidth]{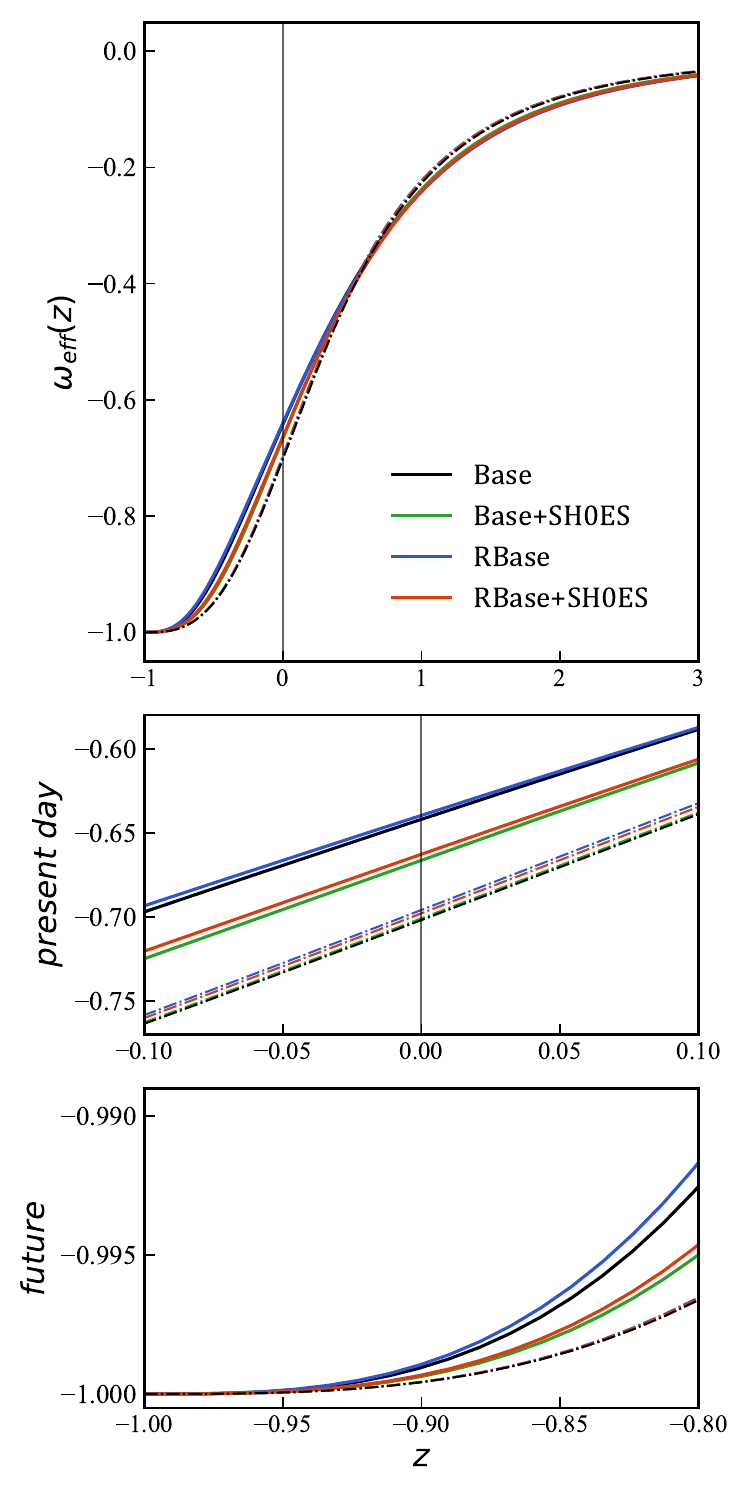}
    \caption{Effective EoS parameter $ \omega_{eff}(z) $. The solid lines represent $ f(Q) $ model while the dot-dashed lines represent $ \Lambda $CDM. Each color corresponds to the same combination of datasets.}
    \label{figure_weff}
\end{figure}

\begin{figure}
    \centering
    \includegraphics[width=0.5\textwidth]{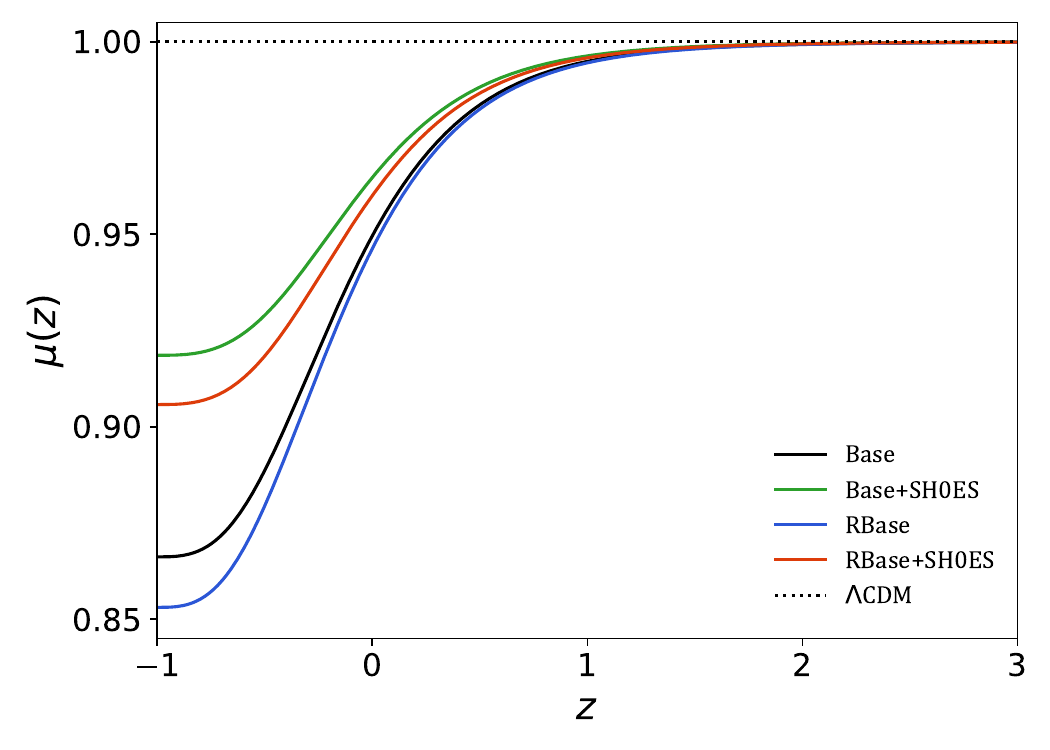}
    \caption{Effective Gravitational coupling $ \mu(z) $. The $ f(Q) $ deviate from $ \mu=1 $ for $ z<2 $ and split into two bands. The (R)Base bands exhibit the most suppressed growth while the SH0ES bands elevate the suppression. We also observe that adding the RSD data causes a drop in suppression amplitude. }
    \label{figure_mu}
\end{figure}

\begin{figure}
    \centering
    \includegraphics[width=0.5\textwidth]{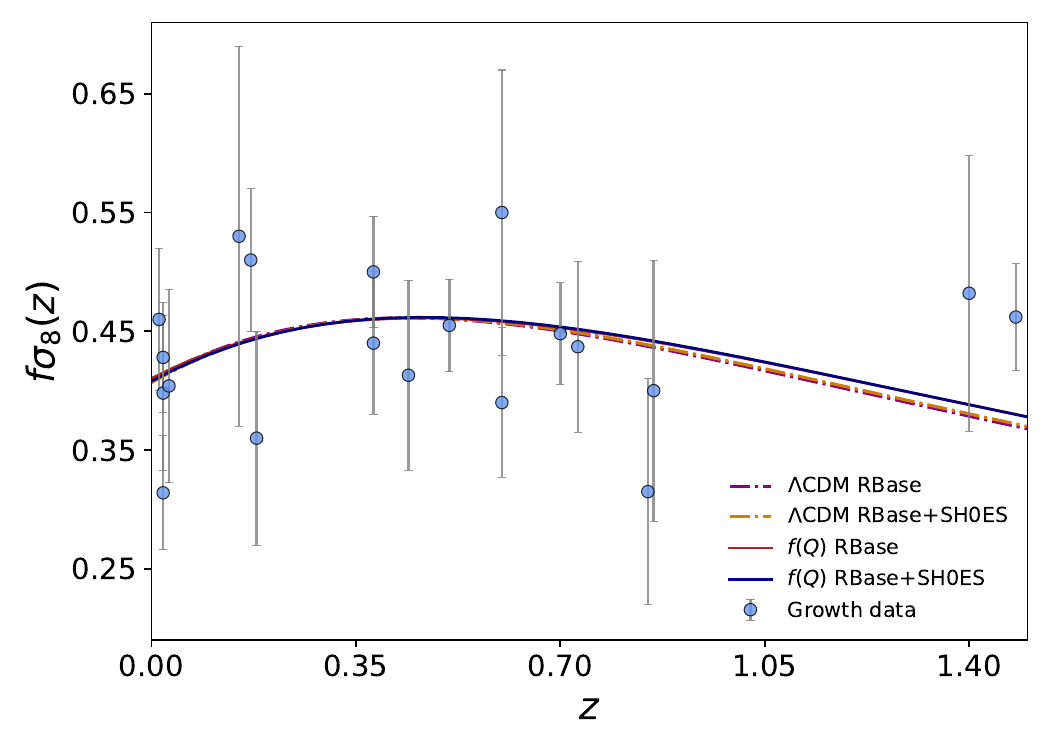}
    \caption{Growth rate $ f\sigma_8(z) $.}
    \label{figure_fs8}
\end{figure}

This subsection reconstructs the evolution of the Universe of the MCMC constrained Hybrid $ f(Q) $ model through fundamental cosmological quantities. \autoref{figure_Hz} shows the reconstructed $ H(z) $ parameter for both models compared to the observed CC data. There is overlapping agreement between the models across all combinations of datasets. 
The deceleration parameter and the effective equation of state parameter are presented in \autoref{figure_deceleration} and \autoref{figure_weff} respectively. The deceleration parameter figure demonstrates how the Hybrid model is equivalent to $ \Lambda $CDM while slightly pulling the transition redshift down by less than $ 3\% $. Regarding the effective EoS parameter, we see that the $ f(Q) $ curves split into two sets of bands, where the SH0ES prior pulls the curves slightly towards $ \Lambda $CDM. Both quantities asymptotically approach $ -1 $ at $ z=-1 $.  
\autoref{figure_mu} displays the evolution of the effective gravitational coupling $ \mu(z) $ effectively confirming our analytical deduction - there is a suppression of gravity under the Hybrid model for all the dataset combinations.
The reconstructed growth rate against the RSD data points is plotted in \autoref{figure_fs8}. The Hybrid model successfully reproduces the $ f\sigma_8 $ velocity field despite the suppressed gravity visible in \autoref{figure_mu}. This proves the kinematic compensation: by inflating the clustering amplitude $ \sigma_{80} $, the Hybrid model correctly compensates its suppressed growth signature, yielding a consistent late-time evolution.
\section{Concluding Remarks and Future}\label{section_conclusion}
In this work, we explored the Hybrid model in the $ f(Q) $ gravity regime and whether it could serve as a viable alternative to the cosmological constant $ \Lambda $. Before injecting the model into the statistical pipeline, we mathematically delineated the regions of viability manifested through constraints on the parameter space. After establishing the positive $ H(z) $ branch as the true expansion history given by the $ f(Q) $ geometry, we reduce the parameter space from $ (\alpha_1, \alpha_2, \alpha_3)\rightarrow(\alpha_1, \alpha_2) $ by imposing the present day condition $ H(z=0) = H_0 $. Subsequently, we find a region of validity in $ \alpha_2\in[0, 2(\alpha_1 - \sqrt{\alpha_1\Omega_0})]\cup[2(\alpha_1 + \sqrt{\alpha_1\Omega_0}), +\infty] $ by demanding the realness of $ H(z) $ at all redshifts. We then recast the modified Friedmann equation into a fluid form of spatially flat + dynamical dark energy and immediately realize the need to set the linear coupling equal to $ 1 $, avoiding any phantom pathologies or early dark energy domination. This in turn forces the Hybrid model to mimic an expansion history identical to $ \Lambda $CDM at the background level.\\

After cementing the mathematical viability of our model, we ran the MCMC analysis on a custom Bayesian inference pipeline, separating the pure background from background + growth data analyses. The Base combination comprising of CC+SN+BAO datasets prefers the $ f(Q) $ model moderately over $ \Lambda $CDM\footnote{We have inferred the Jeffrey's scale as given in \cite{deCruzPerez:2024shj}}, while adding the high-$ H_0 $ SH0ES prior diminishes this preference into the weak regime. A similar trend is observed on adding RSD growth data, although, the information criteria now show a higher preference of the Hybrid model. Across all the dataset combinations, we obtain a negative $ \alpha_3 $ suppressing the gravitational amplitude. Consequently, the model compensates for this weaker gravity by amplifying the late-time clustering amplitude $ \sigma_{80} $ while improving the statistical superiority of the Hybrid model.\\

We also plot the reconstructed dynamical quantities - $ H(z),\;q(z),\;\omega_{eff}(z) $ and $ f\sigma_8(z) $. All the combinations yield a consistent set of observable parameters today, agreeing on the deceleration-acceleration redshift and the quintessence nature of the Universe. We also report on the $ S_8 $ parameter defined by $ S_8 = \sigma_{80}\sqrt{\Omega_{m0}/0.3} $. The compensatory growth is evident in the higher values of $ S_8 $ due to the fact that $ \Omega_{m0} $ remains the same for all dataset combinations.\\

Finally, we have isolated the physical novelty of the Hybrid $ f(Q) $ to the linear perturbation sector. The resulting amplitude compensation mechanism provides a distinct and falsifiable $ S_8 $ signature. While the current RSD data reveal a positive statistical preference of this suppressed gravitational regime, the definitive test of this framework will emerge through confronting it with the full-shape cosmic shear and weak lensing map analysis from both the current and upcoming surveys like DES, LSST and Euclid.

\bibliography{main}

\end{document}